# Giant Nernst effect and field-enhanced transversal $z_N T$ in ZrTe$_5$

Peipei Wang[1,2], Chang-woo Cho[2], Fangdong Tang[2], Peng Wang[2], Wenjie Zhang[3], Mingquan He[4], Genda Gu[5], Xiaosong Wu[3], Yonghong Shao[1*], Liyuan Zhang[2*]

[1] Key Laboratory of Optoelectronic Devices and Systems of Ministry of Education and Guangdong Province, College of Optoelectric Engineering, Shenzhen University, 518060 Shenzhen, China.

[2] Department of Physics, Southern University of Science and Technology, 518055 Shenzhen, China.

3 State Key Laboratory for Artificial Microstructure and Mesoscopic Physics, Beijing Key Laboratory of Quantum Devices, Peking University, 100871 Beijing, China.

4 Low Temperature Physics Lab, College of Physics & Center for Quantum Materials and Devices, Chongqing University, Chongqing 400044, China.

5 Condensed Matter Physics and Materials Science Department, Brookhaven National Laboratory, Upton, NY, USA.

## Abstract

*Thermoelectric materials can recover electrical energy from waste heat and vice versa, which are of great significance in green energy harvesting and solid state refrigerator. The thermoelectric figure of merit (zT) quantifies the energy conversion efficiency, and a large Seebeck or Nernst effect is crucial for the development of thermoelectric devices. Here we present a significantly large Nernst thermopower in topological semimetal ZrTe$_5$, which is attributed to both strong Berry curvature and bipolar transport. The largest in-plane $S_{xy}$ (when **B**//**b**) approaches 1900 µV/K at T=100K and **B**=13T, and the out-of-plane $S_{xz}$ (when **B**//**c**) reaches 5000 µV/K. As a critical part of $z_N T$, the linearly increased in-plane $S_{xy}$ and resistivity $\rho_{yy}$ regard to **B** induces an almost linear increasing transversal $z_N T$ without saturate under high fields. The maximum $z_N T$ of 0.12 was obtained at **B**=13 T and T= 120K, which significantly surmounts its longitudinal counterpart under the same condition.*

* corresponding author: shaoyh@szu.edu.cn, zhangly@sustech.edu.cn

# Introduction

Pursuing high-performance thermoelectric materials and modulatory methods have a long history in thermoelectric applications. In the past few decades, versatile methods and efforts such as engineering band structure[1,2], seeking materials with intrinsic low thermal conductivity or by nanostructuring[3-5], introducing quantum confinement effect[6], fabricating sophisticated heterostructure[7,8] etc. have been attempted to realize high figure of merit ($zT$). The $zT$ had been greatly improved over the generally desired threshold value of 2[4,5,8]. So far, most studies were focusing on the longitudinal $zT$, where the electrical current $I$ is along the heat current $Q$ direction. Compare to the well-studied longitudinal $zT$, the rarely explored transversal thermoelectric devices (where $I \perp Q$) allows arbitrary spatial configurations and separation of $n$- or $p$- type legs are not necessary, thus have more convenience in the potential application. More importantly, the orthogonality between heat and electrical could break the intertwined relationship between the off diagonal thermal conductivity and electrical conductivity governed by the Wiedman-Franz law. This advantage could provide a possibility to tune the thermal conductivity and electrical conductivity independently and achieve an optimized $zT$.

The topological materials (topological insulators, Dirac/Weyl semimetals) have been investigated extensively these years. These studies revoked the interests in the thermoelectric field. Most Dirac/Weyl semimetals possess high mobility and large Fermi velocity, their physical properties are thus readily affected by the application of an external magnetic field. It is suggested that magnetic field could be a novel tuning knob to enhance $zT$ in topological materials[9-11] and the thermopower of 3D Dirac materials can increase unboundedly with field well beyond the quantum limit[11,12,14-16]. Indeed, an order of magnitude enhancement of $zT$ from 0.17 to 1.1 has been realized in Dirac semimetal $Cd_3As_2$ with a field of 7 T[13]. Similarly, large transversal thermopower $S_{xy}$ can also be obtained in Dirac/Weyl semimetals due to strong Berry curvature[17-19] and bipolar transport of compensated electrons and holes[20]. Such field enhanced transversal $z_NT$ has also been found in $Cd_3As_2$, where the $z_NT$ runs up to 0.7 exceeding the longitudinal counterpart above 100 K[21].

Here, we focus on the bulk topological semimetal $ZrTe_5$. $ZrTe_5$ was initially noticed because of the giant resistivity peak[22,23]. The mechanism had not been fully understood, and the ARPES result showed that the chemical potential of $ZrTe_5$ is sensitive to the temperature changes[24]. Meanwhile, the topological properties of $ZrTe_5$ crystals are in debate among strong topological insulator (TI), weak TI and Dirac/Weyl semimetal[25]. Although ambiguous in the classification, it has attracted much attention recently due to various exotic experimental discoveries[26-33]. In this paper, we investigate the transversal magneto-thermoelectrical properties of $ZrTe_5$ at various temperatures. We find a large anomalous Nernst effect both when $B//b$ and $B//c$, the amplitude increases as the chemical potential $\epsilon$ approach the Dirac point, reflecting the impact of underlying Berry curvature. Well beyond the quantum limit regime, the in-plane (***B**//**b***) Nernst thermopower $S_{xy}$ and magnetoresistivity $\rho_{yy}$ increase linearly or quasi-linearly with ***B***. Consequently, a field enhanced $z_N T$ without saturation up to ***B***=13 T is realized experimentally. The maximum value of 0.12 is obtained at *T*=120K and ***B***=13T, with a slope of $d(z_N T)/dB = 0.005/$T.

## Experimental setup

ZrTe$_5$ single crystals were grown by Te flux methods[30]. Resistivity $\rho_{xx}$ and $\rho_{yy}$ were measured using a standard four-probe method. Before making the electrical connections, the crystal surface was cleaned by Argon plasma followed by a deposition of 5/30 nm thick Cr/Au pot pattern. This enables us to make Ohmic contacts with typical contact resistance of ~ 1 Ohm. For the thermal conductivity and thermoelectrical measurements, the thermal current is applied along ***a*** axis, while the magnetic field is parallel to the ***b*** or ***c*** axis. The temperature difference $\Delta T$ was measured using two pairs of calibrated type-E thermocouples, which were attached at the same position of voltage contacts. All the data had been symmetrized in positive and negative fields to exclude the asymmetric effect.

## Results and Discussion

Figure 1a presents the orthorhombic crystal structure of ZrTe$_5$ with space group Cmcm (No.63), the three principal axes *a*, *b* and *c* correspond to the directions *x*, *z* and *y*, respectively. As temperature is cooled down from 300 K, a well-known resistivity peak appears at 90 K ($T_p$) accompanied by a sign change in the Seebeck signal $S_{xx}$ as shown in Fig. 1b. This implies a switching of carrier types from high temperature hole type to low temperature electron type as reported in previous studies[14,27]. A recent theoretical study has shown that the resistivity anomaly is caused by the Dirac polarons, which are mixtures of massive Dirac electrons and holes encircling a cloud of phonons[34]. At temperatures low enough, the Dirac polarons do not influence the transport properties effectively, the system behaves as a Dirac semimetal which possesses ultra-low carrier density ($8*10^{16}$/cm$^3$) and extremely high mobility (over 500,000 cm$^2$/V.s). The low electron density enables the quantum limit to be reached at around 1.5T, above which $\mu B \gg 1$ is satisfied. Fig.1c shows the magnetic field dependent $\rho_{xx}$ and $S_{xx}$ taken at 3.2K. The well-developed quantum oscillations reveal a single electron pocket as small as 1.26T at the Fermi level, as shown in Fig. 1d, indicate the one-band nature of the system. The quantum oscillations can be resolved at about 0.2 T, which corresponds to a large magnetic length $l_B = \sqrt{\hbar/eB} \approx 57$ nm, characterizing the high quality of the sample. The Fermi level is estimated to about 20m*eV* above the Dirac point at $T$=1.5K (see Supplementary Fig. S1).

Now we turn to investigate the field dependence of Nernst $S_{xy}$ at selected temperatures. In Fig. 2a and 2b, we present the in-plane Nernst $S_{xy}$ as a function of the magnetic field. As shown in Fig. 2a and 2b, complex behavior of $S_{xy}$ is observed, which clearly deviates from the semiclassical picture $S_{xy} \propto \mu B/(1+\mu^2 B^2)$ deduced from the Mott relation. In the high field limit $\mu B \gg 1$, varnishing Nernst signal $S_{xy}$ is expected in the semiclassical formula. But surprisingly, here the $S_{xy}$ grows linearly with field and shows no sign of saturation up to 13 T. Similar behavior has been reported in NbP and TaAs[17,20], which is attributed to a strong bipolar contribution to $S_{xy}$ due to symmetric electron-hole excitations. More evidences of bipolar transport are shown in Supplementary Fig. S2 and S4. At low field (below 2T), $S_{xy}$ increase sharply initially and then reach a plateau at around 1-2T. The trend is most pronounced around 100 K. This step-like profile of $S_{xy}$ signifies an anomalous Nernst effect stemming from the underlying Berry curvature when relevant symmetries are broken[17,19,26]. The observed total $S_{xy}$ contains the traditional and anomalous contributions, here we employ the empirical formula[19] to separate these two parts:

$$S_{xy} = S_{xy}^N \frac{\mu B}{1+(\mu B)^2} + S_{xy}^A \tanh \frac{B}{B_0} \qquad (1)$$

Where $\mu$ is the carrier mobility, $B_0$ is the saturation field above which the signal reaches its plateau value, $S_{xy}^N$ and $S_{xy}^A$ are the amplitudes of the conventional and anomalous Nernst signals, respectively. This empirical equation describes our experimental data quite well, as shown in Fig.2e. The fitted anomalous Nernst coefficient $S_{xy}^A$ normalized by temperature $T$ has been plotted in Fig. 2f (black curve).

As $T$ rises, the $S_{xy}^A/T$ increases from 2.4 $\mu V/K^2$ at 30 K to its maximum value of 9.4 $\mu V/K^2$ at 100 K, then drops quickly at higher temperatures. The behavior of $S_{xy}^A/T$ can be explained in the semi-classical Boltzmann scenario, where the anomalous conductivity tensor $\sigma_{xy}^A$ and thermoelectric tensor $\alpha_{xy}^A$ are related to the Berry curvature as[18]:

$$\sigma_{xy}^A = \frac{e^2}{\hbar} \sum_n \int \frac{d^3k}{(2\pi)^3} \Omega_n^z f_{n,\mathbf{k}} \qquad (2)$$

$$\alpha_{xy}^A = \frac{k_B e}{\hbar} \sum_n \int \frac{d^3k}{(2\pi)^3} \Omega_n^z s_{n,\mathbf{k}} \qquad (3)$$

Here $\Omega_n^z$ is the Berry curvature with $n$ a band index, $f_{n,\mathbf{k}} = f(E_{n\mathbf{k}})$ is the Fermi distribution function and $s_{n,\mathbf{k}} = -f_{n,\mathbf{k}} \ln f_{n,\mathbf{k}} - [(1-f_{n,\mathbf{k}})\ln(1-f_{n,\mathbf{k}})]$ the entropy density for the dispersion $E_{n,\mathbf{k}}$ of the conduction electron band $n$. The Nernst coefficient is generally expressed in terms of the thermoelectric tensor $\alpha_{ij}$ and conductivity tensor $\sigma_{ij}$:

$$S_{xy} = \frac{\alpha_{xy}\sigma_{xx} - \alpha_{xx}\sigma_{xy}}{\sigma_{xx}^2 + \sigma_{xy}^2} \quad (4)$$

Eqs. (2) (3) and (4) indicate that the amplitude of anomalous Nernst signal is strongly dependent on the intensity of Berry curvature near the chemical potential, that is, depending on the energy difference between the chemical potential $\epsilon$ and the Dirac point. As $\epsilon$ moves towards the Dirac node, the AHE becomes more and more prominent and slight variation of the $\epsilon$ affects the AHE significantly. For ZrTe$_5$, as $T$ rises the chemical potential continuously sweeps over the Dirac node from conduction band to valence band[24]. It's reasonable that the $S_{xy}/T$ attains its maximum value around $T_p$ where the temperature induced Lifshitz transition occurs. However, the maxima appear at 100K is slightly higher than the resistivity peak temperature 90K. This is possibly due to the peak shift under the high magnetic field[27,30].

We further examine the out-of-plane Nernst thermopower $S_{xz}$ when **B**//**c**, as shown in Fig. 2c and 2d. The tendency of $S_{xz}$ is similar to the low field region of the $S_{xy}$, but the amplitude of 5000 $\mu V/K$ at 100K is significantly large, which can be addressed as 'giant' Nernst effect. The main contribution comes from the anomalous part, we use Eq. (1) to fit the $S_{xz}$ and shows the results in Fig. 2f (red curve). The anomalous amplitude $S_{xz}/T$ increases with temperature, reach its maxima of 36 $\mu V/K^2$ at 100K. Above 100K, the fittings are diverging. Despite the high anisotropy of thermoelectric properties for **B**//**c** and **B**//**b**, similar temperature and field dependencies of the anomalous Nernst coefficient $S^A/T$ along both directions point to the same origin, i.e., the Berry curvature.

The observed large and non-saturating field induced transversal thermoelectric response motivated us to investigate the transversal figure of merit of ZrTe$_5$, which is defined as $z_N T = \frac{S_{xy}^2 T}{\rho_{yy}\kappa_{xx}}$. Fig. 3a-d present the corresponding in-plane resistivity $\rho_{yy}$, thermal conductivity $\kappa_{xx}$, power factor $pF_{tr}$ (which is $S_{xy}^2/\rho_{yy}$), and $z_N T$ respectively, as a function of field **B** at selected temperatures. For the magnetoresistivity $\rho_{yy}$, it increases quasi-linearly or linearly with **B** above 4 T. Upon increasing the temperature from 30 K, the high field value of $\rho_{yy}$ decreases. This is quite different from that of $\rho_{xx}$ (see Supplementary Fig. S3a) which shows quadratic behavior at temperatures away from $T_p$ with a maximum value appears at 110 K. Fig. 3b shows that the thermal conductivity is less sensitive to the **B**-field. The magnitude of $\kappa_{xx}$ is slightly suppressed at moderate field and eventually becomes nearly a constant above 4 T, which reflects the mixed component of electrical thermal conductivity $\kappa_{xx}^e$ and the lattice thermal conductivity $\kappa_{xx}^l$. We can roughly estimate the $\kappa_{xx}^e$ using WF law by assuming the Sommerfeld value $L_0 =$

$\kappa_{xx}^e \rho_{xx}/T = 2.45 * 10^{-8}$ WΩK$^{-2}$. Simple calculation shows that that $\kappa_{xx}^e$ accounts for 4% (at 0 T) and 0.2% (at 13 T) of the total $\kappa_{xx}$. This is reasonable since the carrier density is low and becomes even lower when the temperature approaches $T_p$. Above the quantum limit, $\kappa_{xx}^e$ is completely suppressed leaving out the lattice part $\kappa_{xx}^l$ only which is essentially independent of **B**. While $\kappa_{xx}$ remains constant, both quantities $S_{xy}$ and $\rho_{yy}$ increase linearly in the high field regime. Therefore, both the transversal power factor $pF_{tr}$ and $Z_NT$ are enhanced by the field, as shown in Fig. 3c and 3d. Both quantities climb monotonically with **B** above 4 T and no sign of saturation is found up to 13 T. The maximal in-plane transversal figure of merit up to 0.12 is obtained at *T*=120K and **B**=13T. We also present the longitudinal thermoelectric properties in Supplementary Fig. S3d, where the longitudinal thermopower $S_{xx}$ increases linearly without saturate under high magnetic field, consistent with the theoretical predictions in Ref. 11 12. The longitudinal *zT* have a relatively larger value at zero field but are quickly suppressed by the magnetic field. The field enhancement of *zT* only appears at around 110K, but the values are much smaller than the $z_NT$ due to the relatively small value of the thermopower $S_{xx}$.

Despite the marked linear increase profile under high fields, the small $z_NT$ value of 0.12 is not likely useful for practical thermoelectric applications. Compare to Cd$_3$As$_2$ where a higher $z_NT$ of 0.7 was obtained[21], the large magnetoresistivity $\rho_{yy}$ of ZrTe$_5$ (110 mΩ.cm at 9T and 50K, which is about 50 times larger than that of Cd$_3$As$_2$) prevents the realization of a sizable transversal $z_NT$. A simple strategy is that we can adjust the carrier density by doping or intercalation to enhance the electrical conductivity. However, this may cause a reduction of thermopower and an enhancement of electrical thermal conductivity, which are unfriendly for obtaining high performance. But in the case of ZrTe$_5$, the temperature tunable chemical potential makes it easy to tune the intensity of Berry curvature. When the chemical potential lies in the vicinity of Dirac point, large Nernst signals are expected both from the anomalous Nernst effect and the bipolar transport even if the doping level is high. Consequently, a large field-induced power factor $pF_{tr}$ is expected if the carrier density lies in an appropriate range. Moreover, as ZrTe$_5$ has a relatively low thermal conductivity and the magnetic field can suppress it further, it is hopeful to realize field-induced high $z_NT$ in this system after proper band structure engineering. It is worth noting that ZrTe$_5$ has a giant out-of-plane Nernst thermopower $S_{xz}$ over 5 mV/K when **B**//*c*, tuning transport parameters in this direction may lead to excellent thermoelectric performance.

## Conclusion

In conclusion, we have studied the field induced transversal thermoelectric properties of the topological semimetal ZrTe$_5$. Under conditions that chemical potential

$\epsilon$ is close to the Dirac point and magnetic field is high (when *H//b*), the Nernst thermopower $S_{xy}$ is significantly enhanced to 1900 $\mu V/K$ (at *T*=100K and *B*=13T). Such a field induced enhancement of Nernst effect is benefited from both the anomalous Nernst effect arising from the pronounced Berry curvature and the bipolar transport of compensated electrons and holes. Consequently, linear enhancement and non-saturating transversal figure of merit $z_N T$ up to 13 T is realized in this system. Our findings thus open a new avenue towards high-efficient transverse thermoelectricity.

## Acknowledgements

The authors thank Liang Fu and Brian Skinner for their helpful discussion. This work was supported by Guangdong Innovative and Entrepreneurial Research Team Program (No. 2016ZT06D348), NFSC (11874193), China postdoctoral Science Foundation (2020M672760), and Shenzhen Fundamental subject research Program (JCYJ20170817110751776). M. He acknowledges the support by National Natural Science Foundation of China (11904040), Chongqing Research Program of Basic Research and Frontier Technology, China (Grant No. cstc2020jcyj-msxmX0263), Fundamental Research Funds for the Central Universities, China (2020CDJQY-A056, 2020CDJ-LHZZ-010), Projects of President Foundation of Chongqing University, China(2019CDXZWL002).

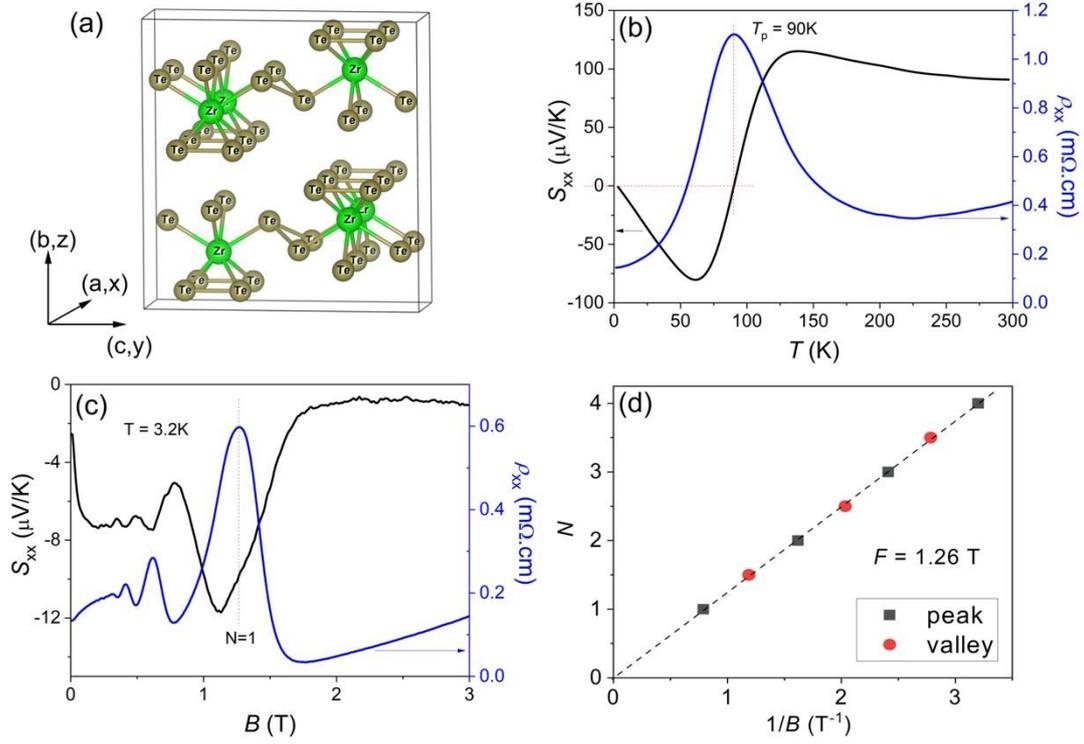

**Figure 1** **(a)** Crystal structure of ZrTe5. **(b)** Temperature dependence of the Seebeck $S_{xx}$ (black curve) and resistivity $\rho_{xx}$ (blue curve) under zero field. **(c)** $S_{xx}$ and $\rho_{xx}$ as a function of field B taken at 2K, both show clear quantum oscillations. The red dashed line marks the N=1 Landau level, above which $\rho_{xx}$ decreases sharply, then enters the quantum limit above 1.5T. **(d)** Landau fan diagram of the Landau level index $N$ versus 1/B at 3.2K. Black and red dot indicate the peak and valley positions, respectively.

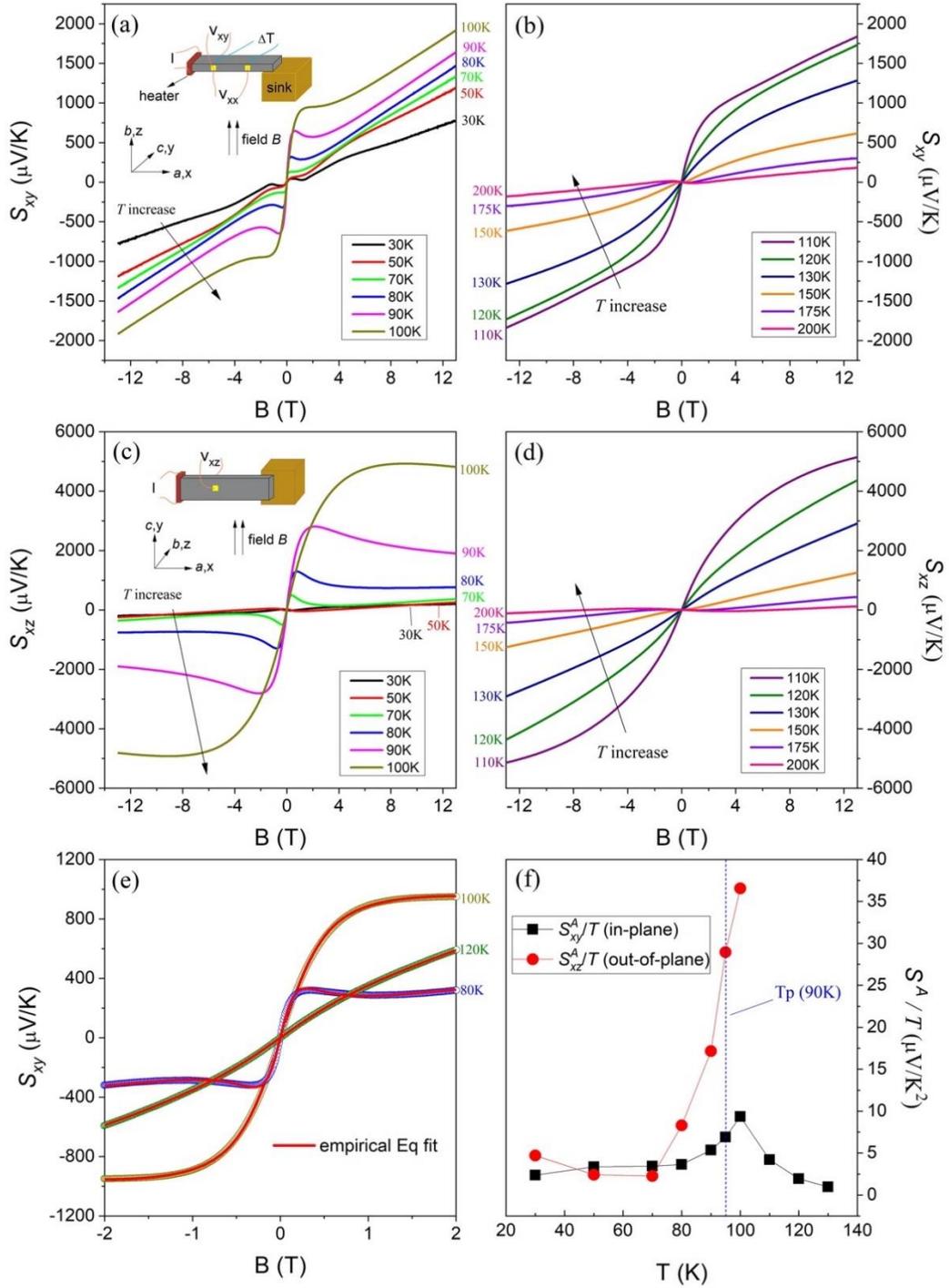

**Figure 2**. **(a)(b)** The in-plane Nernst thermopower $S_{xy}$ as a function of $B$ under different temperatures, the field $B$ is parallel to the $b$ axis. The inset shows the schematic thermoelectric measurements setup of $S_{xy}$. **(c)(d)** The out-of-plane Nernst $S_{xz}$ as a function of field $B$ under different temperatures, where the field is parallel to the $c$ axis. The inset shows the schematic thermoelectric measurements setup of $S_{xz}$. **(e)** Fit the anomalous Nernst signals of $S_{xy}$ from -2T to 2T under selected temperatures using Eq.1. **(f)** The fitted anomalous Nernst coefficient $S^A/T$ as a function of temperature $T$. Black and red lines indicate the in-plane $S^A_{xy}/T$ and out-of-plane $S^A_{xz}/T$, respectively. The dashed line indicates the resistivity peak temperature 90K.

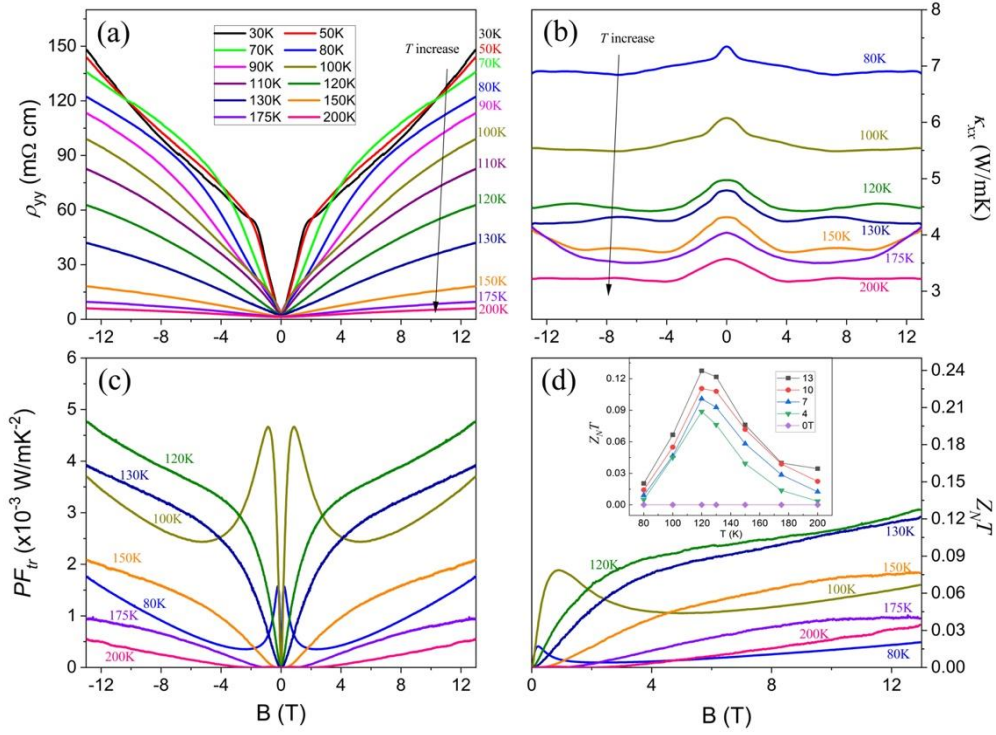

**Figure 3** **(a)** The magnetoresistivity $\rho_{yy}$ at different temperature $T$. **(b)** The thermal conductivity $\kappa_{xx}$ taken from 80K to 270K. As temperature rises, the $\kappa_{xx}$ decreases monotonically. **(c)** The field dependence of power factor $pF_{tr}$. **(d)** The transversal $z_N T$ as a function of field ***B***. Generally, high field could enhance the $z_N T$, the greatest enhancement occurs at 120K. Inset: plot the $z_N T$ as a function of temperature $T$ under fixed magnetic field ranging from 0 to 13T.

# Supplementary Information for

# Large Nernst effect and field enhanced transversal ZT in ZrTe$_5$

Peipei Wang[1,2], Chang-woo Cho[2], Fangdong Tang[2], Peng Wang[2], Wenjie Zhang[3], Genda Gu[4], Xiaosong Wu[3], Yonghong Shao[1*], Liyuan Zhang[2*]

1. Key Laboratory of Optoelectronic Devices and Systems of Ministry of Education and Guangdong Province, College of Optoelectric Engineering, Shenzhen University, 518060 Shenzhen, China.
2. Department of Physics, Southern University of Science and Technology, 518055 Shenzhen, China.
3. State Key Laboratory for Artificial Microstructure and Mesoscopic Physics, Beijing Key Laboratory of Quantum Devices, Peking University, 100871 Beijing, China.
4. Condensed Matter Physics and Materials Science Department, Brookhaven National Laboratory, Upton, NY, USA.

## 1. Analyze the quantum oscillations

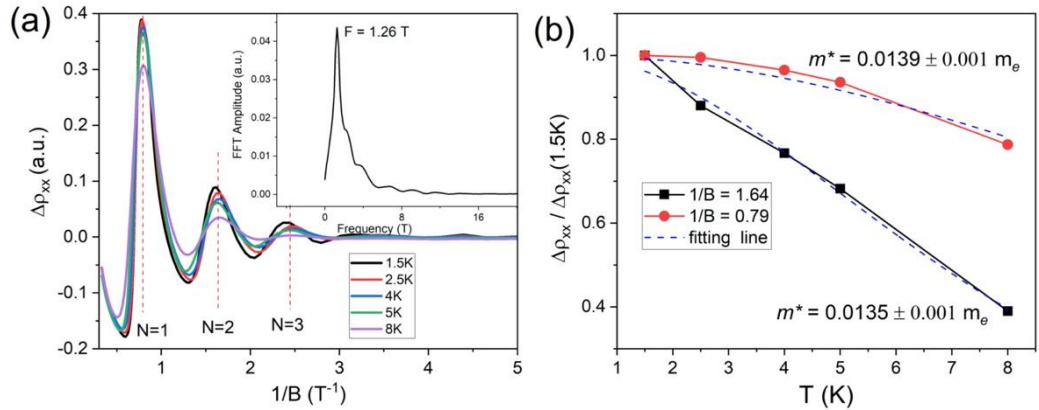

**Figure S4** (a) The longitudinal resistivity $\rho_{xx}$ (when $B$ // b axis) as a function of 1/B after subtracting the non-oscillating background, under selected temperatures. The red dashed lines indicate the first three Landau levels. Inset shows a typical FFT spectrum of the 1.5K data, which reveals a single dominating frequency of 1.26 T. (b) Normalized SdH oscillation amplitude as a function of temperature, black and red lines are taken at $B$ = 0.61T and $B$ =1.26T respectively. The blue dashed fitting line based on the temperature damping factor $R_T = \alpha X/\sinh(\alpha X)$ ($\alpha = 2\pi^2 k_B/e\hbar$; $X = m^*T/B$) yields $m^*$ = 0.014 m$_e$.

The low temperature resistivity $\rho_{xx}$ of ZrTe5 shows clear Shubnikov-de Haas (SdH) quantum oscillations, they are periodic in 1/$B$ as shown in Fig. s1a. The FFT analysis reveals a single frequency of 1.26T. With this frequency, the Fermi surface cross section area $A_F$ can be estmimated using the Onsager relation $F = A_F\hbar/2\pi e$. Consequently, the $k_F$ can be calculated from $A_F = \pi k_F^2$ if we assume a circular Fermi surface cross section. By analyzing the temperature dependence of the SdH oscillation amplitude following the Lifshitz-Kosevich theory (Figure S1**b**), we can extract the effective cyclotron mass $m^*$. Given the values of $k_F$ and $m^*$, the Fermi velocity $v_F = \hbar k_F/m^*$ and the Fermi energy $\epsilon_F = \hbar v_F k_F$ can be calculated. The obtained parameters are: cross section area $A_F = 1.2 \times 10^{-4}$ Å$^{-2}$, Fermi wave vetor $k_F = 0.62 \times 10^{-2}$ Å$^{-1}$, cyclotron mass $m^* = 0.014 m_e$, Fermi velocity $v_F = 5.1 \times 10^6$ $m/s$, and the Fermi energy $\epsilon_F = \hbar v_F k_F = 19.7\ meV$ above the Dirac point at 1.5K.

## 2. Temperature dependence of the carrier density and mobility

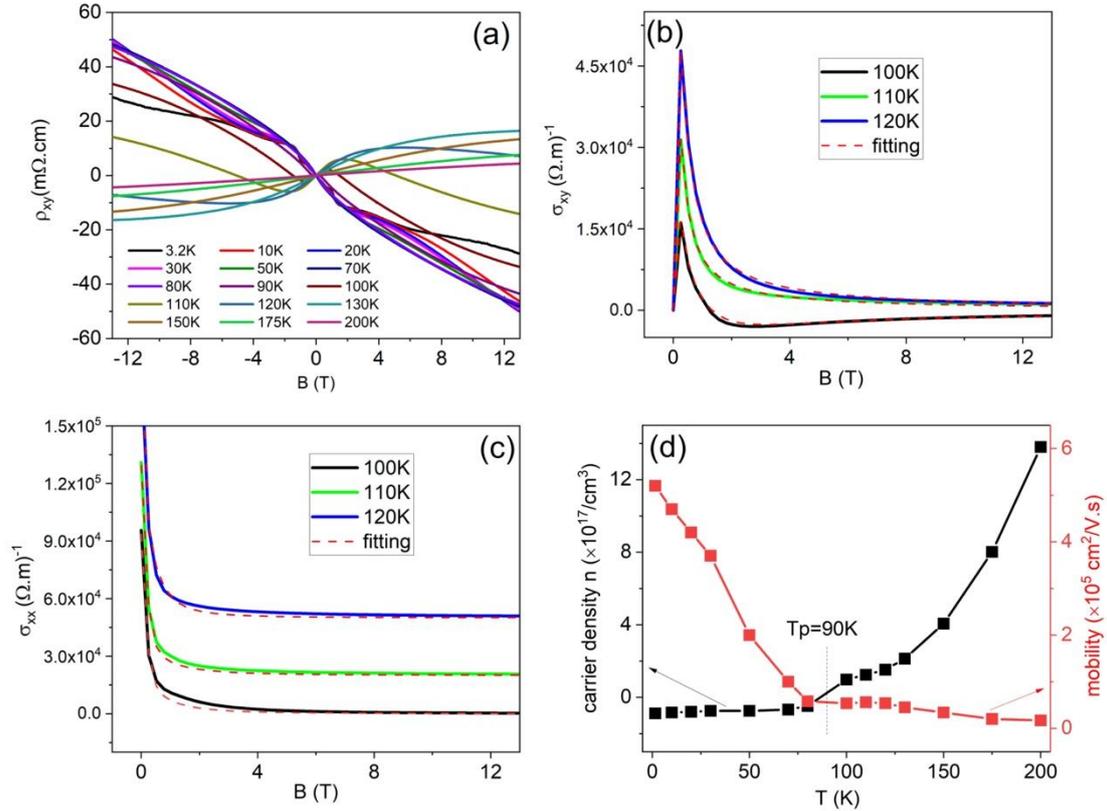

**Figure s5** (a) The Hall resistivity $\rho_{xy}$ under different temperatures when $B//\mathbf{b}$ axis. (b) Two-band model fitting of the $\sigma_{xy}$ using Eq. S1 at T=100K, 110K, 120K. (c) Two-band model fitting of the $\sigma_{xx}$. The curves have been shifted for clarity. (d) The fitting yield the temperature dependence of the carrier density **n** (black curve) and the mobility $\mu$ (red curve).

In Figure S2**a**, we show the Hall resistivity $\rho_{xy}$ at selected temperatures; for the corresponding curves of $\rho_{xx}$, see Figure S4**a**. At temperatures that are much lower or much higher than 90 K, the Fermi level is not so close to the Dirac point and the transport is more or less one-carrier dominated at low field. Correspondingly, the low-field Hall resistivity is relatively linear and the carrier density can be simply calculated from $\rho_{xy} = B/ne$, and the mobility can be calculated from $\sigma = ne\mu$. On the other hand, around 90 K, where the Fermi level is in the vicinity of the Dirac point, a significant number of electrons and holes are thermally excited. The existence of two types of carriers causes non-linear behavior of $\rho_{xy}$ even at low field. To describe the magnetotransport around $T_p$, we employ the two-band model here:

$$\sigma_{xx}(B) = \frac{n_e e \mu_e}{1+\mu_e^2 B^2} + \frac{n_h e \mu_h}{1+\mu_h^2 B^2} \ ; \quad \sigma_{xy}(B_z) = \frac{n_e e \mu_e^2 B}{1+\mu_e^2 B^2} + \frac{n_h e \mu_h^2 B}{1+\mu_h^2 B^2}. \qquad (S1)$$

With

$$\sigma_{xx} = \frac{\rho_{xx}}{\rho_{xx}^2 + \rho_{xy}^2} \ ; \quad \sigma_{xy} = \frac{\rho_{xy}}{\rho_{xx}^2 + \rho_{xy}^2}. \qquad (S2)$$

In Figure S2(b) and (c), the red dashed line indicate the fitting of the experimental data $\sigma_{xy}$ and $\sigma_{xx}$ at 100K, 110K and 120K, the fittings are carried out simultaneously to minimize the error. We can see that the Eq.S1 describe the experimental data quite well. The fitting yields the carrier density n and the mobility $\mu$ of the dominant carrier under different temperatures, which are shown in Fig. S2d. A transition from electron to hole dominated transport is clearly seen around temperature $T_p$.

## 3. Longitudinal *ZT*

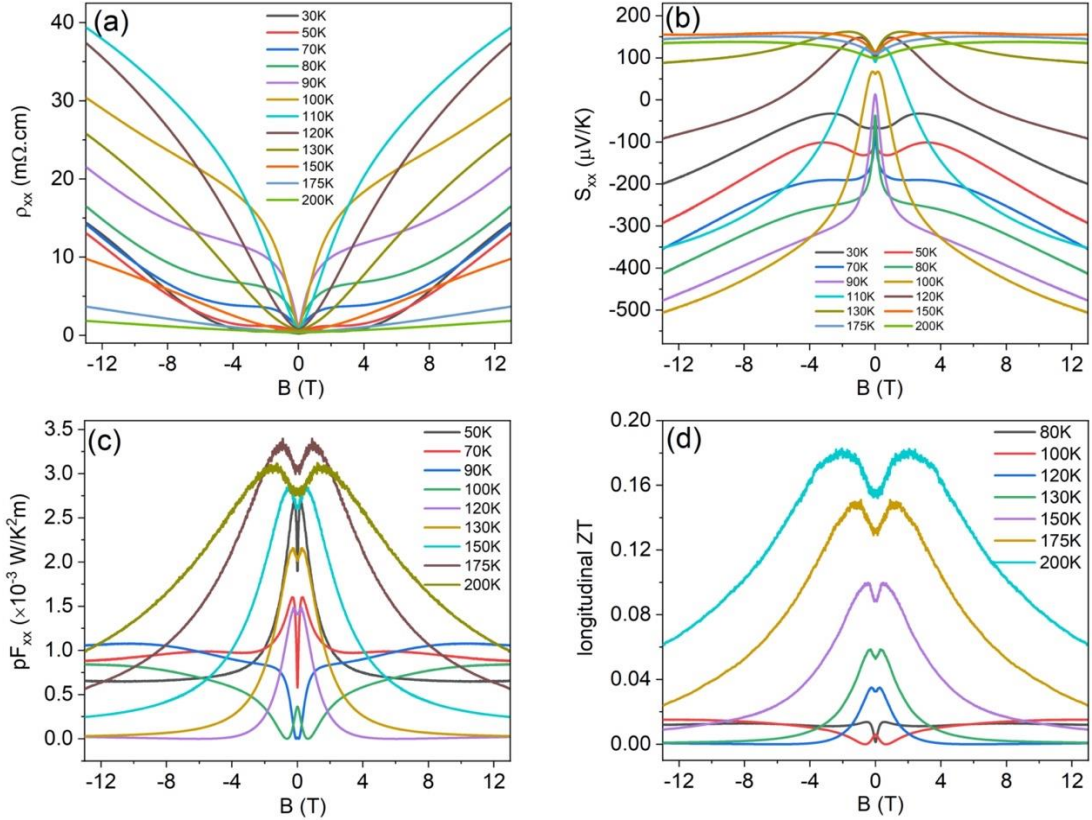

**Figure S6** (a) The longitudinal resistivity $\rho_{xx}$ as a function of field *B* under different temperatures when *B*//**b** axis. (b) The thermopower $S_{xx}$ as a function of field *B* under different temperatures when *B*//**b** axis. (c) The field dependence of longitudinal power factor pF (which is $S_{xx}^2/\rho_{xx}$) at selected temperatures. (d) With the obtained longitudinal power factor pF and thermal conductivity $\kappa_{xx}$ (see Fig. 3b in the main text), the longitudinal

$ZT$ (which is $S_{xx}^2 T/\kappa_{xx}\rho_{xx}$) are displayed.

The longitudinal magnetoresistivity $\rho_{xx}$ (Fig. S3a) and the magnetothermopower $S_{xx}$ (Fig. S3b) when **B** is parallels to **b** axis are similar as previously reported data [Ref 13, 25]. Compare to the $\rho_{yy}$, the $\rho_{xx}$ shows anomalous behavior at around 110K, where it attains the largest value and the behavior is more or less linear regard to **B**. At higher or lower temperatures, the high field behavior becomes quadratic. This is disadvantage for realizing high *pF* or *ZT*. Fig. S3b present the thermopower $S_{xx}$ at selected temperatures, it increase linearly with **B** above the quantum limit and shows no sign of saturating. However, this linearly increasing is compensated by the quadratic increasing of $\rho_{xx}$, finally results in a suppression of *pF* and longitudinal *ZT* with field **B** as shown in Fig. S3c and S3d. The only exception happens at around 110K because of the non-quadratic behavior of $\rho_{xx}$. But due to the relatively small value of $S_{xx}$ (500 $\mu V/K$ under 13T field, which is 3 times smaller than the corresponding $S_{xy}$) and the rapid increasing of $\rho_{xx}$ (40 mΩ.cm under 13T field), the obtained longitudinal $ZT$ is less than 0.02 at around 120K, which is 5 times smaller than its transversal counterpart $Z_N T$.

As pointed out in Ref 11, three effects could lead to a linearly increase of Sxx in a 3D Dirac/Weyl semimetal with a linear dispersion and gapless band structure. First, a sufficiently high magnetic field produces a large enhancement of the electronic density of states and a reduction in the Fermi energy $E_F$. Second, a quantizing magnetic field assures that the transverse $E \times B$ drift of carriers plays a dominant role in the charge transport, and this allows both electrons and holes to contribute additively to the thermopower. Third, in materials with a small band gap and electron-hole symmetry, the Fermi level remains close to the band edge in the limit of large magnetic field, and this allows the numbers of thermally excited electrons and holes to grow with magnetic field even while their difference remains fixed.

In the case of Dirac semimetal ZrTe5, it have a linear dispersion and a small band gap which are required by the theoretical prediction, and an extremely high mobility and ultra-low carrier density which enables us to reach the quantum limit already at 1.5T. This small value of quantum limit makes it perfect to investigate the quantum limit physics. We have present the field induced longitudinal Seebeck coefficient $S_{xx}$ in Fig. 1c and Fig. S3b. One can see that below the quantum limit 1.5T, the $S_{xx}$ present clear quantum oscillations, and well beyond the quantum limit, the $S_{xx}$ increase linearly with the field *B* without saturate. These features are precisely the same as predicted in Ref.11.

## 4. One-band fits to $S_{xx}$ and $S_{xy}$

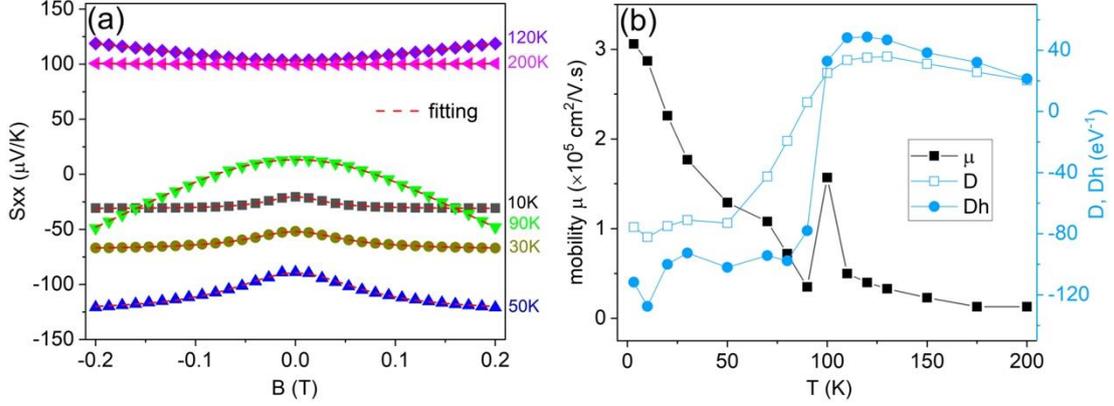

**Figure S7** (a) The fitting of the week-$B$ thermopower $S_{xx}$ data using equation S3 (red dashed line). (b) The obtained parameters $\mu$, $D$ and $D_H$ from the fitting as a function of temperature $T$.

As shown in ref. 15, in the semiclassical regime, the field dependence of thermopower $S_{xx}$ and Nernst $S_{xy}$ can be expressed in the form of the electrical conductivity $\sigma_{xx}$ and $\sigma_{xy}$ in a single band system:

$$S_{xx} = A \left( \frac{\sigma_{xx}^2}{\sigma_{xx}^2+\sigma_{xy}^2} D + \frac{\sigma_{xy}^2}{\sigma_{xx}^2+\sigma_{xy}^2} D_H \right) \tag{S3}$$

$$S_{xy} = A \frac{\sigma_{xx}\sigma_{xy}}{\sigma_{xx}^2+\sigma_{xy}^2} (D_H - D) \tag{S4}$$

Where $A = \frac{\pi^2 k_B^2 T}{3e}$, and $D = \partial\ln\sigma_{xx}/\partial\varepsilon$ and $D_H = \partial\ln\sigma_{xy}/\partial\varepsilon$ are the longitudinal and transverse differential logarithmic conductivities at chamical potencial $\varepsilon_F$, respectively. Fig. S4a present the fitting of the experimental low field data $S_{xx}$ using equation S3 from 10K to 200K. We can see that the above equation describes the measured data quite well in the whole temperature range. The obtained parameters $\mu$, $D$ and $D_H$ under each temperature are plotted in Fig. S4b. Compare to the resistivity fitting results in Fig. S2d, the independent fitting of $S_{xx}$ yields a similar value of the mobility $\mu$ and a similar tendency about temperature $T$. The difference happens at 100K-120K, where the thermal excitation of holes participate in the transport process, leading to the breakdown of the one-band assumption. This bipolar transport will also result in a discrepancy of the parameter $D_H$ -$D$ obtained from the independent fitting of $S_{xx}$ and $S_{xy}$ respectively, as shown in ref. 10 and 15. However, in the case of ZrTe$_5$, the fitting of $S_{xy}$ using Eq. S4 failed because of the existence of anomalous Nernst effect, which is very different from the formular S4 both in the sign and the behavior regard to $B$.